
\input phyzzx

\Pubnum={{\vbox {QMW 92/14 \hbox {hep-th/yymmnn}}}}
\pubtype={}
\date={August 1992}

\titlepage

\title{The geometry of supersymmetric coset models and superconformal
algebras  }

\author{G. Papadopoulos  \foot {Address from October 1st: Department of
Mathematics, King's College London, Strand, London WC2R 2LS.}}

\address{Department of Physics,\break Queen Mary and Westfield College,\break
              Mile End Road,\break London E1 4NS, UK.}

\abstract{An on-shell formulation of $(p,q)$, $2\leq p\leq 4$, $0\leq q\leq
4$, supersymmetric coset models with target space the group $G$ and gauge
group a subgroup $H$ of $G$ is given.  It is shown that there is a
correspondence between the number of supersymmetries of a coset model and the
geometry of the coset space $G/H$.  The algebras of currents of supersymmetric
coset models are superconformal algebras.  In particular, the algebras of
currents of (2,2) and (4,0) supersymmetric coset models are related to the
N=2 Kazama-Suzuki and  N=4 Van Proeyen superconformal algebras
correspondingly.}

\endpage

\pagenumber=2


\def\fff{\vrule width0.5pt height5pt depth1pt}
\def\pp{{{ =\hskip-3.75pt{\fff}}\hskip3.75pt }}

\def \CL {\cal {L}}
\def \CN {\cal {N}}
\def \CM {\cal {M}}
\def \CJ {\cal {J}}
\def \CE {\cal {E}}


\REF\zumino {B. Zumino, Phys. Lett.  \underbar{87B} (1979), 205.}
\REF\alvarez {L. Alvarez-Gaum\' e and D.Z. Freedman, Commun. Math. Phys.
\underbar {80} (1981), 443.}
\REF\hull{S.J. Gates, C.M. Hull and M. Rocek, Nucl. Phys. \underbar {B248}
(1984), 157.}
\REF\hulla{C.M. Hull, in \lq\lq Super Field Theories", ed. H.C. Lee et al
(Plenum, New York, 1987).}
\REF\ga {P.S. Howe and G. Papadopoulos, Nucl. Phys. \underbar {B289} (1987),
264.}
\REF\gb {P.S. Howe and G. Papadopoulos, Class. Quantum Grav. \underbar{5}
(1988), 1647.}
\REF\hps  {C.M. Hull, G. Papadopoulos and B. Spence,
												Nucl. Phys. \underbar {B363} (1991), 593.}
\REF\witten {E. Witten, Commun. Math. Phys. \underbar{92} (1984), 455.}
\REF\gaugeda{
M. Douglas, CalTech preprint CALT 86-1453 (1987);\hfill\break
             E. Guadagnini, M. Martellini and M. Minchev, Phys. Lett.
													\underbar {191B}  (1987), 69;\hfill\break
              W. Nahm, Davis preprint UCD 88-02 (1988); \hfill\break
             K. Bardacki,
              E. Rabinovici and B. S\"aring, Nucl. Phys. \underbar{B299}
											  (1988), 157;\hfill\break
              P. Bowcock, Nucl. Phys. \underbar{B316}  (1989), 80; \hfill\break
             H.J. Schnitzer, Nucl. Phys. \underbar{B324} (1989),412;
												\hfill\break
             D. Karabali and H.J. Schnitzer, Nucl. Phys.
												\underbar{B329} (1990), 649; \hfill\break
             K. Gaw\c edzki and A. Kupiainen, Phys. Lett. \underbar {215B}
									  (1988), 119 and Nucl. Phys. \underbar{B320}(FS) (1989),
											625;\hfill\break
             J.M. Figueroa-O'Farrill, Stony Brook preprint ITP-SB-89-41}
\REF\gaugedb {C. M. Hull and B. Spence, Mod.  Phys. Lett. \underbar{A6} (1991),
969 and Nucl Phys. \underbar {B353} (1991), 379; \hfill\break
I. Jack, D.R.T Jones, N. Mohammedi and H. Osborn, Nucl. Phys. \underbar {B332}
(1990), 359.}
\REF\gko  {P. Goddard, A. Kent
and D. Olive, Phys. Lett. \underbar {152B} (1985), 88; Commun. Math. Phys.
\underbar{103} (1986), 105; \underbar {113} (1987),1.}
 \REF \ks {Y. Kazama and H. Suzuki, Phys. Lett. \underbar {216B} (1989),
             112;  Nucl. Phys. \underbar {B321} (1989), 232.}
\REF\hs {C.M. Hull and
B. Spence, Phys. Lett. \underbar {241B} (1990), 357. }
\REF\nak {T. Nakatsu, \lq\lq Supersymmetric gauged Wess-Zumino-Witten models",
\hfill \break
preprint UT-587 July-1991.}
\REF\vp {A. Van Proeyen, Class. Quantum Grav.
\underbar {6} (1989), 1501. }  \REF\hsone{C.M. Hull and B. Spence, Phys. Lett.
\underbar {232B} (1989), 204.}  \REF\gp{G.Papadopoulos, Phys. Lett. \underbar
{277B} (1992), 447.} \REF\svp {Ph. Spindel, A. Sevrin, W. Troost and A. Van
Proeyen, Nucl. Phys. \underbar {B308} (1988), 662; \underbar {B311} (1988),
465.}   \REF\fn {A. Frohlicher and A. Nijenhuis, Nederl. Wetensch. Proc. Ser. A
\underbar {59} (1956), 339.}
\REF\zam {A.B. Zamolodchikov, Teor. Mat. \underbar {65} (1985), 1205.}
\REF\ish  {S. Ishihara, J. Differential Geom. \underbar {9} (1974), 483. }
\REF\sa  {S. Salamon, Invent. Math. \underbar {67} (1982), 143.}
\REF\wolf {J. Wolf, J. Math. Mech. \underbar{14} (1965), 1033.}


\chapter {Introduction}

It has been known for many years that there is an interplay between the
geometry
of the target manifolds of supersymmetric sigma models and their number of
supersymmetries [\zumino].   In two dimensions, sigma models may have left-
and/or right- handed supersymmetries and the geometry of their target
manifolds has been extensively studied in the literature [\alvarez-\gb].

More recently, the authors of Ref [\hps] studied the two-dimensional
supersymmetric {\it gauged} sigma models and found that there is a similar
interplay between their number of supersymmetries and the geometry of their
target manifolds.  The same authors introduced off-shell multiplets and
actions for all $(p,q)$, $2\leq p \leq 4$, $0\leq q\leq 2$, supersymmetric
gauged sigma models by generalising a method employed in Refs  [\ga,\gb] for
the
construction of off-shell multiplets and actions for the (ungauged)
two-dimensional supersymmetric sigma models.

An interesting class of (supersymmetric) gauged sigma models arises from
gauging the (supersymmetric) Wess-Zumino-Witten (WZW) models
[\witten].  These models have been studied by many authors [\gaugeda,\gaugedb],
they are (super)conformal and their  algebras of currents realise
(super)conformal algebras that are constructed from a method devised by
Goddard, Kent and Olive known as the GKO or coset construction [\gko].  In the
following, we will refer to gauged WZW  models as coset models.

In Ref [\ks]  Kazama and
Suzuki (KS) constructed an N=2 superconformal algebra from an N=1
superconformal algebra using the GKO construction.   The KS algebra may
have a left and a right sector.  A realisation of the left or right sector of
the KS algebra was given in Ref [\hs] in terms of the currents of an off-shell
(2,0) or (0,2) supersymmetric coset model.
The expectation is that the KS superconformal algebra with both left and right
sectors is realised as the algebra of currents of the (2,2) supersymmetric
coset model.  Several attempts were made to construct this model (see for
example Ref [\nak]) including the  off-shell formulation of Ref [\hps].
However
the  off-shell  closure of the supersymmetry algebra imposes strong conditions
on the geometry of the sigma model target manifold.  In the case of (2,2)
supersymmetric coset model, these conditions restrict the geometry of the group
manifold in such a way that the algebra of currents of these models is not the
most general realisation of the KS algebra.

A. Van Proeyen (VP) in Ref [\vp] constructed an
non-linear left-handed  N=4 superconformal algebra using the GKO
construction and the algebraic properties of tensors on Lie
algebras that are associated with  symmetric quaternionic K\"ahler manifolds
(Wolf spaces).  It is expected that this algebra can be realised as the
algebra of the currents of the (4,0) supersymmetric coset model.  However as in
the case of the (2,2) supersymmetric coset model, the algebra of currents of
the
off-shell (4,0) model constructed in Ref [\hps] does not provide the most
general realisation of the VP superconformal algebra.

In this paper, we will construct an {\it {on-shell}} formulation of all
$(p,q)$,
$2\leq p\leq 4$, $0\leq q \leq 4$ supersymmetric coset models with target
manifold  the group $G$ and gauge group a subgroup $H$ of $G$.  We
will achieve this by starting with the (1,1) or (1,0) supersymmetric coset
models and then by constructing the additional supersymmetry transformations
necessary for the description of $(p,q)$ supersymmetric coset models.   Then
we will derive the conditions  for the action of (1,1) or
(1,0) coset model to be  invariant under the action of $(p,q)$ supersymmetry
transformations and give the conditions  for the on-shell closure of the
algebra of $(p,q)$ supersymmetry transformations.  We will find that
these conditions have a geometric interpretation.  Indeed they will be
understood in terms of the geometry of the coset space $G/H$.   In particular
we will prove that (2,q), $0\leq q \leq 2$, models exist provided that $G/H$ is
a Hermitian manifold with a holomorphic tangent bundle and (4,$q$), $0\leq q
\leq 4$ models exist provided that $G/H$ is a quaternionic K\"ahler manifold
with respect to its canonical connection. The algebras of currents of
supersymmetric coset models are superconformal algebras.  Finally, the
algebraic
closure properties of the algebra of supersymmetry transformations, the current
content and the geometry of the coset spaces $G/H$ of the (2,2) and (4,0)
supersymmetric coset models indicate that the algebras of currents of these
models are (classical) realisations of the N=2 KS and N=4 VP superconformal
algebras correspondingly.

This paper has been organised as follows:  In section two, we will present
the (1,1) and (1,0) supersymmetric gauged models and set up our notation.
In section three, we will give the (2,2) supersymmetry transformations  and
calculate their commutator.  In section four, we will derive the conditions
necessary for the on-shell closure of the algebra of the supersymmetry
transformations of a (2,2) supersymmetric coset model and relate them to the
geometry of its target manifold.  In addition, we will examine briefly the
(2,0) and (2,1) supersymmetric coset models.  In section five, we will
construct
the (4,0) supersymmetric coset model and discuss the geometry of the
associated coset space $G/H$.  Finally in section six, we will describe the
rest
of the $(p,q)$ models and give our conclusions.

\chapter {The (1,1) and (1,0) supersymmetric coset models}

The action of the (1,1) supersymmetric gauged sigma model with a
Wess-Zumino term  [\hsone] is
$$\eqalign {
 L =&\  g_{ij} \nabla _{+} X^i\  \nabla _{-} X^j
+ b_{ij}\  D_{+}X^i\  D_{-} X^j -
\cr
&- A^a_{+}\  u_{ia}\  D_- X^i- A^a_{-}\  u_{ia}\
D_{+} X^i + c_{ab}\ A^a_{-}\  A^b_{+}	}
																																																																\eqn\aone$$
where $b_{ij}=-b_{ji}$, $X$ is a section of a bundle with base space an (1,1)
superspace with co-ordinates $(y^{\pp}, y^=, \theta^-, \theta^+)$ and fiber a
manifold $M$, $i,j,k=1, \cdots, {\rm{dim}}{ }M$, and
$\{D_{+},D_{-},\partial_{\pp},\partial_{=}\}$ are the flat superspace
derivatives ($D^2_{+}= i \partial_{\pp}$, $D^2 _{-}= i \partial_{=}$).  The
target space $M$ admits a group action of a gauge group $H$ which generates the
vector fields $\xi_a$, $a=1,\cdots, {\rm{dim}}{ }LieH$, where $LieH$  is the
Lie algebra of the group $H$. The metric $g$ and the three form $B= 3db$ of
$M$ are invariant tensors under the group action of $H$ on $M$.  $\xi_a$ are
Killing vector fields,
$$ \xi^i_a B_{ijk}=\  2\
\partial_{[j} u_{k]a}           \eqn\atwo$$
 and
$$c_{ab} = \xi^i_a u_{ib}.																															\eqn\athree$$
$\{A_{+},A_{-},A_{\pp},A_{=}\}$ are the components of a  connection $A$ and
$\{\nabla _{+},\nabla _{-},\nabla _{\pp},\nabla _{=}\}$ are the corresponding
covariant derivatives that satisfy the (super)algebra
 $$\eqalign{
[\nabla _{+},\nabla _{-}]&=F_{+-},\  [\nabla
_{-},\nabla _{\pp}]=F_{-\pp}, \cr
 [\nabla _{+},\nabla _{=}]&= F_{+=},\  [\nabla _{\pp},\nabla _{=}]=F_{\pp=},
\cr
[\nabla _{+},\nabla _{+}]&= 2 i \nabla_{\pp},\  [\nabla _{-},\nabla _{-}]= 2 i
\nabla_{=} .}
																																																															\eqn\afour$$
   The rest of the supercommutators vanish and  $F$ is the curvature
of the connection $A$.  The action \aone\ is gauge invariant provided that
${\CL}_a u_b = {f_{ab}}^c u_c$ and $c_{ab}=c_{[ab]}$ where ${f_{ab}}^c$ are the
structure constants of the Lie algebra $LieH$ and ${\CL}_a$ is the Lie
derivative of the vector field $\xi_a$.   Finally, $\nabla_+X^i=D_+X^i + A^a_+
\xi^i_a$  and $\nabla_-X^i=D_-X^i + A^a_- \xi^i_a$.

In the case of coset models, the target manifold $M$ is a group  $G$
and $H$ is a subgroup of $G$.  The group action of $H$ on $G$ is
$k\rightarrow hkh^{-1}$ where $k\in G$ and $h\in H$.  This is the adjoint
action
of $H$ into $G$.  The metric $g$ is chosen to be a  bi-invariant positive
definite metric on $G$, i.e. $g_{ij}=L^A_i L^B_j \delta_{AB}=R^A_i R^B_j
\delta_{AB}$, $A,B=1,\cdots, {\rm {dim}}{ }LieG$.  $k^{-1} dk= L^A t_A$ and $dk
k^{-1} =R^A t_A$ , $k\in G$, are the left and right frames correspondingly
where
$t_A$ is an orthonormal basis of the Lie algebra $LieG$ of $G$.  $B$ is a
bi-invariant three form on $G$ ($B_{ijk}=-L^A_iL^B_jL^C_k
f_{ABC}=-R^A_iR^B_jR^C_k f_{ABC}$).  The Killing vector fields are  given by
$\xi_a=L_a -R_a$ and $u_a=L_a +R_a$. $LieH$ is a subspace of $LieG$ and we have
decomposed $LieG$ into $LieH$ and its orthogonal complement, i.e.
$LieG=LieH\oplus P$.  Under this decomposition, $LieG$ is reductive, i.e.
$[LieH, P]\subset P$.  Moreover due to a relative normalisation of the
Wess-Zumino  and the Kinetic terms of the theory, the connections
$$ {\Gamma ^{(\pm) i}}_{jk}\  = \Gamma ^{i}_{jk}(g)
\pm {1\over 2} {B^{i}}_{jk}
\eqn\afive$$
are flat where $\Gamma (g)$ is the Levi-Civita connection of
the bi-invariant metric $g$.

The equations of motion of an (1,1) supersymmetric coset model are the
following:

$$ \nabla^{(+)}_- \nabla_+ X^i -  L^i_a  F^a_{+-} = 0,
																																																														\eqn\asix$$
$$ L_{ia}\  \nabla _+ X^i=0																								\eqn\aseven$$
and
$$ R_{ia}\  \nabla _- X^i=0,																							\eqn\aeight$$
where $\nabla^{(\pm)}$ are the covariant derivatives of the $\Gamma^{(\pm)}$
connections.  A consequence of eqns. \asix\ and \aseven\ is that the
curvature $F$ of the connection $A$ is zero on-shell.  This is a key  property
of the (1,1) ((1,0)) supersymmetric coset models  that allows us to
construct the additional supersymmetry transformations in order to
describe the $(p,q)$, $2\leq p\leq 4$, $0\leq q \leq 4$ models from the
(1,1) (or (1,0)) ones and prove the on-shell closure of the associated
supersymmetry algebras.

We conclude this section with a brief description of the (1,0)
supersymmetric coset model.   Let $\xi_a$ be the vector fields generated by
the the group action of a group $H$ on the target manifold $M$ of an (1,0)
supersymmetric sigma model with Wess-Zumino term $b$ ($b_{ij}=-b_{ji}$).  The
action of the gauged sigma model with gauge group $H$  is
$$\eqalign {
 L =&\  g_{ij} \nabla _{+} X^i\  \nabla _{=} X^j
+ b_{ij}\  D_{+}X^i\  \partial_{=} X^j -
\cr
&- A^a_{+}\  u_{ia}\  \partial_= X^i+ A^a_{=}\  u_{ia}\
D_{+} X^i - c_{ab}\ A^a_{=}\  A^b_{+},	}
																																																																\eqn\aanine$$
where $X$ is a section of bundle with base space the (1,0) superspace with
co-ordinates $(y^{\pp}, y^=, \theta^+)$ and fibre the target manifold $M$,
$\{A_+, A_=, A_{\pp}\}$ are the components of the gauge connection $A$ and $\{
\nabla_+, \nabla_=, \nabla_{\pp}\}$ are the corresponding covariant derivatives
that satisfy the (super)algebra $$[\nabla_+,\nabla_+]=2\ i\ \nabla_{\pp},\quad
[\nabla_+, \nabla_=]=F_{+=},  \quad
[\nabla_=,\nabla_{\pp}]=F_{=\pp}.																 \eqn\aaten$$ $F$ is the
curvature of the connection $A$.  $D_+$ is the flat superspace derivative,
$D_+^2= i \partial_{\pp}$.   The rest of the notation is the same as in the
(1,1) model and the action \aanine\ is gauge invariant provided that ${\CL}_a
u_b = {f_{ab}}^c u_c$ and $c_{ab}=c_{[ab]}$.

To describe the (1,0) coset models, we take as a target manifold $M$ to be a
group $G$. The gauge group $H$ is a subgroup of $G$ that acts on $G$
with the adjoint action as in the (1,1) coset model.  Similarly we choose the
metric $g$ and the Wess-Zumino term $B=3db$ of the model.   The equations of
motion of the (1,0) coset model are $$ \nabla^{(+)}_= \nabla_+ X^i -  L^i_a
F^a_{+=} = 0,
																																																														\eqn\aaeleven$$
$$ L_{ia}\  \nabla _+ X^i=0																								\eqn\aatwelve$$
and
$$ R_{ia}\  \nabla _= X^i=0.																							\eqn\aathirteen$$
A consequence of these equations of motion is that $F=0$ (on-shell).


\chapter {The (2,2) coset model}

To describe the (2,2) supersymmetric coset model, we begin with the (1,1)
supersymmetric coset model and then introduce the additional supersymmetry
transformations necessary for the construction of this model.  The (2,2)
supersymmetric coset model has two left-handed and two right-handed
supersymmetry transformations.  In the following, we concentrate on the
left-handed symmetries of the action \aone.  At the end of this
section, we will return to the right-handed symmetries and give the commutator
of the left-handed transformations with the right- handed ones.  It is
straightforward to calculate the commutator of the right-handed transformations
from the commutator of the left-handed ones.

The action \aone\ is manifestly (1,1) supersymmetric. The first left-handed
supersymmetry and translation transformations are
$$\eqalign{
\delta_L X^i&= \eta \nabla _{\pp}X^i -{i\over 2} D_+\eta \nabla_+X^i
\cr
\delta_L A^a_- &=-i \eta \nabla_+F^a_{+-} + {i\over 2} D_+\eta F^a_{+-}
\cr
\delta_L A^a_+ &=0 }                           \eqn\anine$$
where $\eta=\eta(x^{\pp}, \theta^+)$ are the parameters of the
transformations.

The second left-handed supersymmetry transformation
can be written as follows:
$$ \delta_L X^i = a_-\ {{I}^i}_{j}\ \nabla_+X^{j},						\eqn\aten$$
and
$$\eqalign { \delta_L A^a_- &= a_-\  {D^a}_b \ F^b_{+-};    \cr
\delta_L A^a_+ &=0.}																								\eqn\aeleven$$
$I$ is an {\it{invariant}} and $D$ is an {\it{equivariant}} tensor of $G$
under the adjoint  action of the group $H$ in $G$ and $a_-=a_-(x^{\pp},
\theta^+)$ is the parameter of the transformation.    The transformations
\aten\ and \aeleven\ are a special case of the most general higher spin
transformations given in Ref [\gp].  The invariance property of $I$ will be
suitably modified in section five for the construction of (4,0) supersymmetric
coset models . The transformations \aten\ and \aeleven\  are symmetries of
the action \aone\  provided that
$$\nabla ^{(+)}_k {I^i}_{j}=0,
																																											\eqn\athirteen$$
$$I_{ij}=I_{[ij]}
																																											\eqn\afourteen$$
and
$$ L^k_a   I_{k i}=- {D^b}_a\,
L_{bi}.
																																														\eqn\afifteen$$
To find the condition \athirteen, we have used the fact that $I$ is an
invariant tensor under the adjoint action of $H$ on $G$.   The commutator of
two left-handed supersymmetry transformations on the field $X$ is
$$ [\delta_L, \delta'_L] X^i= \delta ^{(1)} X^i + \delta ^{(2)} X^i+
\delta ^{(3)} X^i																																												\eqn\asixteen$$
where
$$ \delta ^{(1)} X^i=
  a_-'\  a_-  {\CN } (I)^i_{jk} \
\nabla_+X^j  \nabla_+X^k,           \eqn\aseventeen$$

 $$\delta ^{(2)} X^i  =
-\ D_+(a_-' a_-)\
\  {I^i}_{k}\ {I^k}_{j}\  \nabla_+X^{j}												\eqn\aeighteen$$

and

$$ \delta ^{(3)} X^i =-\ 2 i a_-' a_-  {I^i}_{k}\ {I^k}_{j} \nabla_{\pp}X^{j}
\eqn\anineteen $$
 where ${\CN }(I)$ is the Nijenhuis tensor [\fn] of the tensor $I$ and is
given by
$$
{{\CN }(I)^i}_{jk}= 2 \big{(} {I^l}_{[j} \partial_{|l|} {I^i}_{k]} -
\partial_{[j} {I^l}_{k]} {I^i}_l \big{)}.
   														 \eqn\atwenty$$

The commutator \asixteen\ of two left-handed supersymmetry transformations
(eqn. \aten) is similar to the commutator of two left-handed
supersymmetry transformations of an ungauged N=1 supersymmetric sigma model.
Indeed, the latter can be recovered from \asixteen, if we replace in the eqns.
\aseventeen -\anineteen\ the covariant superspace derivatives $\nabla_+$
with partial superspace ones.

The commutator of two left-handed supersymmetry transformations   on the
field $A$ (eqn. \aeleven) is always proportional to the curvature $F_{+-}$
of the connection $A$ and its covariant derivatives.  This can be verified by a
short explicit calculation.  Since one of the on-shell conditions of the (1,1)
supersymmetric coset models is $F_{+-}=0$, the transformation of eqn.
\aeleven\ closes on-shell.

The right-handed supersymmetry
transformations are
$$ \delta_R X^i = a_{+}\  {{J}^i}_{j}\ \nabla_-X^{j},			\eqn\atwentyone$$
and
$$\eqalign { \delta_R A^a_- &= a_{+}\  {E^a}_b \ F^b_{+-};
\cr
\delta_R A^a_+ &=0,}																																								\eqn\atwentytwo$$
where $a_+ =a_+(x^{=}, \theta^-)$ is a parameter of the transformation.
$J$ is an {\it{invariant}} tensor and $E$ is an {\it{equivariant}} tensor in
the
manifold $G$ under the action of the group $H$  as in the case of the
left-handed transformations.  The conditions for the invariance of the action
 \aone\  under the transformations \atwentyone\ and \atwentytwo\ are  given by
eqns. \afourteen\ and \afifteen, if we set $J$ and $E$ in  the place of $I$ and
$D$ correspondingly, and $J$ is covariantly constant with respect to
$\nabla^{(-)}$ connection, i.e.
$$ \nabla^{(-)}_k {J^i}_j=0. 																\eqn\atwentythree$$
The commutator of two right-handed supersymmetry transformations on the
field $X$  is  similar as the commutator of two left-handed
transformations. The commutator of two right-handed supersymmetry
transformations on the
field $A$ closes on-shell. Finally the commutator of a left-handed with a
right-handed supersymmetry transformation on the field $X$ is
$$[\delta_L,\delta_R]X^i=   \delta^{(1)}X^i + \delta^{(2)}X^i
																																	+\delta^{(3)}X^i
																																						\eqn\atwentyfour$$
where
$$\delta^{(1)}X^i = a_-\  a_+\  {{\CM} (I,J) ^i}_{jk} \nabla_+X^j\
\nabla_-X^k,
																																															\eqn\atwentyfive$$
$$\eqalign {
\delta^{(2)}X^i =&- a_-\  a_+\  ({J^i}_k {I^k}_j - {I^i}_k {J^k}_j)
																															\nabla_- \nabla_+X^j
\cr &
+\ a_-\  a_+\  {I^i}_k {J^k}_j F^a_{+-} \xi^j_a } \eqn\atwentysix$$
and
$$\delta^{(3)}X^i = a_-\  a_+\ ({D^a}_b {J^i}_j + {E^a}_b {I^i}_j ) F^b_{+-}
\xi^j_a. 																								\eqn\atwentyseven$$
The tensor $\CM$ can be expressed in terms of $I$ and $J$ as follows:
$$ {{\CM}^i}_{jk} (I,J)= \nabla_l{J^i}_k\  {I^l}_j + {J^i}_l\  \nabla_k
{I^l}_j - \nabla_l{I^i}_j\  {J^l}_k - {I^i}_l\
\nabla_j{J^l}_k,														\eqn\atwentyeight$$
where
$\nabla$ is the covariant derivative of the  Levi-Civita connection of the
metric $g$ of the group $G$.  Finally, the commutator of a left-handed with a
right-handed transformation on the field $A$ closes on-shell.


\chapter {The geometry of the (2,2) coset model}

In the previous section, we calculated the commutator of the
(2,2) supersymmetry transformations necessary for the description of the (2,2)
supersymmetric coset model.  In this section, we will derive the conditions for
the existence of (2,2) models and show that these conditions have a geometric
interpretation in terms of the geometry of the coset space $G/H$.

First we begin with the commutator of left-handed with right-handed
transformations (eqn. \atwentyfour ).  The terms in this commutator
proportional to the curvature tensor $F$ are zero on-shell since the
vanishing of $F$ is one of the on-shell conditions.  The rest of the
commutator vanishes on-shell as well.  Indeed,  the tensor $\CM$ can be
expressed in terms of $I$, $J$ and the Wess-Zumino term $B$ of the coset model
using the fact that $I$ ($J$) is covariantly constant with respect to
$\nabla^{(+)}$ ($\nabla^{(-)}$) covariant derivative.  Then combining
$\delta^{(1)}$ (eqn. \atwentyfive ) and $\delta^{(2)}$ (eqn. \atwentysix) in
the
commutator \atwentyfour, we get
$$[\delta_L,\delta_R]X^i= - a_-\  a_+\  ({J^i}_k {I^k}_j
- {I^i}_k {J^k}_j)
																															\nabla^{(+)}_- \nabla_+X^j
																																								\eqn\bone$$
up to terms proportional to the curvature $F$.  However $\nabla^{(+)}_-
\nabla_+X^j$ vanishes on-shell (eqn. \asix ).  Consequently the commutator
\bone\ closes on-shell. In conclusion the commutator of extended left-handed
with right-handed supersymmetry transformations closes on-shell without any
restrictions on the geometry of the target manifold $G$ of the (1,1)
supersymmetric coset model.

Next we turn to study the conditions for the invariance of the action under
both left- and right-handed supersymmetry transformations.  The eqns.
\athirteen\ and \atwentythree\ can be solved as in the case of (ungauged) WZW
models with extended supersymmetries.  Indeed, following Refs [\svp] the most
general solution of these equations is
$$ {I^i}_j= L^i_A {I^A}_B L^B_j \qquad {J^i}_j= R^i_A {J^A}_B R^B_j
																																		\eqn\btwo$$
where $({I^A}_B)$ and $({J^A}_B)$ are constant matrices.  Then equation
\afourteen\ can be solved provided that $(I_{AB})$ and $(J_{AB})$ are
antisymmetric matrices.  Before we examine condition \afifteen, we recall
that both $I$ and $J$ are {\it{invariant}} tensors under the adjoint action of
the group $H$ on $G$.  This invariance property imposes the conditions
$$\eqalign{{f_{a C}}^A  \ {I^C}_{B} - {f_{a B}}^C  \ {I^A}_{C}&=0
\cr
{f_{a C}}^A  \ {J^C}_{B} - {f_{a B}}^C  \ {J^A}_{C}&=0}
																																									\eqn\bthree$$
The solutions of \bthree\ that impose the weakest restriction on the
geometry of the target manifold $G$ and the gauge group $H$ of the (1,1)
supersymmetric coset model are
$$({I^A}_{B})|_P=({I^n}_{m}) \qquad
({J^A}_{B})|_P=({J^n}_{m})
																																										\eqn\bfour$$
and all the other components are zero, where the Lie algebra indices $A, B$
of $LieG$ are split according to the vector space  decomposition
$LieG=LieH\oplus P$, i.e. $a=,\cdots,{\rm {dim}}{ }LieH$ and
$n,m=1,\cdots, {\rm {dim}}{ }P$ (${\rm {dim}}{ }P ={\rm {dim}}{ }LieG -{\rm
{dim}}{ }LieH $).  The equation \bthree\ then becomes
$$\eqalign{{f_{a p}}^m  \ {I^p}_{n} - {f_{a n}}^p  \ {I^m}_{p}&=0
\cr
{f_{a p}}^m  \ {J^p}_{n} - {f_{a n}}^p  \ {J^m}_{p}&=0}
																																													\eqn\bbfour$$
If the tensors $I$ and $J$ are chosen as in eqn. \bfour, we can solve
equation \afifteen\ by setting $D=E=0$.  This concludes the study
of the conditions for the invariance of the action \aone\ under (2,2)
supersymmetry transformations.

To study the conditions that a commutator (eqn. \asixteen ) of two left-handed
supersymmetry transformations closes on-shell, we write  the
Nijenhuis tensor ${\CN} ( I)$ of $I$ using eqn. \btwo\ as
$${{\CN} (I)^i}_{jk} = L^i_A\  {{\CN} (I)^A}_{BC}\  L^B_j\  L^C_k.
																																								\eqn\bfive$$
The components of the Nijenhuis tensor ${\CN} (I)$ that contribute to the
commutator \asixteen\  are ${{\CN }(I)^A}_{mn}$.  The rest are proportional
to the equations of motion \aseven\ and they do not affect the on-shell
closure of the algebra.  Finally we can show using eqn. \bfour\ that the
commutator \asixteen\ closes on-shell to (1,1) supersymmetry transformations
(eqn. \anine )  provided that
$$ {{\CN } (I)^p}_{mn}=0, \qquad {I^n}_{p} {I^p}_{m}= - \delta ^n_m
																																					\eqn\bsix$$
and
$${f_{lp}}^a {I^l}_m {I^p}_n - {f_{mn}}^a =0,					\eqn\bseven$$
where $l,m,n,p=1,\cdots, {\rm {dim}}{ }P$ and $a=1,\cdots, {\rm {dim}}{
}LieH$.  The same analysis can be done for the right-handed
supersymmetry transformations.

To find the geometric interpretation of the conditions \btwo, \bfour, \bbfour,
\bsix\ and \bseven\ in terms of the geometry of the coset space $G/H$, we
introduce a local section $s$ of the principal bundle $H\rightarrow
G\rightarrow G/H$; $H$ acts on $G$ from the right, i.e. $k\rightarrow kh$
where $k\in G$ and $h\in H$.  The $LieG$-algebra valued one-form $s^{-1}ds$ on
$G/H$ can be decomposed as follows: $$s^{-1}ds=e^m t_m + \Omega^a
t_a.													 \eqn\beight$$ $e$ is a frame  and $\Omega$ is the canonical
connection of the coset space $G/H$.   The curvature $F$ of the canonical
 connection is $F^a_{\mu \nu} = - {f_{mn}}^a e^m _{\mu} e^n _{\nu}$ and
its torson $T$ is ${T^{\kappa}}_{\mu \nu}= -{1\over 2}
e_p^{\kappa}{f_{mn}}^p e^m_{\mu} e^n _{\nu}$   where $\mu, \nu, \kappa =1,
\cdots , {\rm {dim}}{ }G/H$.    Using the frame $e$ of the coset space, we
define the metric $g_{\mu \nu}= e^m_{\mu} e^n_{\nu} \delta_{mn}$ in $G/H$ and
the tensor ${I^{\mu}}_{\nu} = e^n_{\nu} {I^m}_n e^{\mu}_m$  where ${I^m}_n$ is
given in eqn. \bfour.   The geometric interpretation of eqn. \bbfour\ is that
the tensor ${I^{\mu}}_{\nu}$ is covariantly constant with respect to the
canonical connection of the coset space $G/H$.  This is equivalent to the
fact that ${I^{\mu}}_{\nu}$ is a $G$-invariant tensor on the coset.  From eqn.
\bsix\ we deduce that the tensor ${I^{\mu}}_{\nu}$ is an integrable
complex structure on the coset space $G/H$ and since $(I_{mn})$ is
antisymmetric matrix the metric $g_{\mu \nu}$ is an (1,1) tensor with respect
to this complex structure, i.e. $G/H$ is a Hermitian manifold. Finally from
eqn.
\bseven\ we find that the curvature $F$ of the canonical connection $\Omega$ is
(1,1) $LieH$-valued two-form with respect to the complex structure
${I^{\mu}}_{\nu}$.   The latter implies that the tangent bundle of $G/H$ is a
holomorphic vector bundle.  Conversly, given a coset space $G/H$ which is a
complex manifold with respect to a G-invariant complex structure such that the
G-invariant metric in $G/H$ and the curvature of its canonical connection  are
(1,1) tensors, we can construct a (2,2)  supersymmetric coset model with
gauge group $H$.

An interesting class of (2,2) supersymmetric coset models arises whenever
the target manifold $G$ is a {\it {semisimple}} Lie group and $G/H$ is a
{\it {symmetric}} space.  In this case the torsion $T$ of the canonical
connection is zero and this connection becomes the Levi-Civita connection of
the
G-invariant metric on $G/H$ and the first condition of eqn. \bbfour\ implies
eqn. \bseven.  For symmetric coset spaces, ${{\CN }(I)^p}_{mn}$ is identically
zero and $I$ is covariantly constant with respect to the Levi-Civita connection
of the invariant metric of the symmetric space, i.e. G/H becomes a K\"ahler
manifold.

Given a complex structure $I$ on $G/H$ necessary for the existence of the
second left-handed supersymmetry transformation, it is straightforward to
construct a complex structure $J$ necessary for the construction of a
second right-handed supersymmetry transformation.   Indeed we may set
$J^i_j=R^i_A I^A_B R^B_j$.  Because of this the (2,1) supersymmetric coset
model is in fact invariant under (2,2) supersymmetry
transformations.

It is straightforward to use the results of the previous section to
 treat the (2,0) supersymmetric coset models
\foot{ Our description of the (2,0) supersymmetric coset model is different
from
the one presented by the authors of Ref [\hs].  One of the differencies is
in the choice of the tensor $I$.}.
To do this we start from the (1,0)
supersymmetric coset model and then introduce the additional left-handed
supersymmetry transformations as in eqns. \aten\ and \aeleven.  The conditions
for invariance of the action \aanine\ under the transformations \aten\ and
\aeleven\ are given by eqns. \athirteen -\afifteen.  The commutator of two
left-handed transformations is given in eqn.  \asixteen.  Finally the closure
propoperties of two left-handed supersymmetry transformations are the same as
the closure properties of two left-handed  supersymmetry transformations of the
(2,2) coset model. Consequently the geometry of the coset space $G/H$ of the
(2,0) supersymmetric coset model is the same as the geometry of the
corresponding coset space of the (2,2) model.


\chapter {The geometry of the (4,0) coset model}

The discussion of sections three and four can be extended to describe the
(4,0)  supersymmetric coset model as well.  This can be achieved by introducing
three complex structures on the coset space $G/H$ one for each additional
supersymmetry.  One of the conditions for the closure of the supersymmetry
algebra is that the three complex structures that generate the left-handed
supersymmetries satisfy the algebra of imaginary unit quaternions.  In addition
each complex structure  must obey the condition \bbfour.  Thus all three
complex structures are covariantly constant with respect to the canonical
connection of the coset space $G/H$.  The latter is a very strong condition on
the geometry of the coset manifolds.  In particular, it excludes all the cases
where $G$ is semisimple and $G/H$ is a symmetric space and consequently these
models are not suitable for the realisation of the VP superconformal algebra.
In the following we will describe a (4,0) model that has geometric
properties similar to those of the VP superconformal algebra and at
the conclusions we will comment how this method may be used to study the
rest of the (4,$q$) models.

 To describe the (4,0) supersymmetric coset models, we begin
from the off-shell formulation of (1,0) supersymmetric coset models of
section two and introduce the additional left-handed
supersymmetry transformations
 $$ \eqalign {\delta_L X^i &= a^r_-\ {{I_r}^i}_{j}\
\nabla_+X^{j} \cr					 \delta_L A^a_- &= a^r_-\  {D_r^a}_b \ F^b_{+-};    \cr
\delta_L A^a_+ &=0; \quad r=1,2,3,}							\eqn\cfour$$
where $a^r_-$ are the parameters of the transformations and $I_r$ are
{\it{equivariant}} tensor fields on $G$, i.e. ${\CL}_aI_r = -{\omega
_{ar}}^sI_s$, $r,s=1,2,3$. $\omega$ is a representation of $LieH$ on the space
spanned by $I_r$ and ${\CL}_a$ is the Lie derivative of of the vector field
$\xi_a$ on $G$. The transformations \cfour\ are symmetries of the (1,0)
supersymmetric coset model provided that the eqns.\athirteen, \afourteen\ and
\afifteen\ are satisfied for every pair of $(I_r, D_r)$ and the parameters
$a^r_-$ are {\it {covariantly}} constant with respect to $\nabla_=$ covariant
derivative, i.e. $$\nabla_=a^r_-=\partial_= a^r_- + A^b_= {\omega_{b s}}^r
a^s_-=0.  \eqn\cfive$$
 From the conditions for the invariance  of the action, each tensor $I_r$
may be written as
 $$ {(I_r)^i}_j= L^i_A {(I_r)^A}_B L^B_j \
																																											\eqn\csix$$
where $({(I_r)^A}_B)$ are constant antisymmetric
matrices ($I_{rAB}=-I_{rBA}$).  As in the case of (2,2) supersymmetric coset
model in section three, we decompose $LieG$ into $LieH \oplus P$ and choose
$({(I_r)^A}_B)|_P=({(I_r)^n}_m)$ with the rest of the components to vanish.
Then we may set $D_r=0$ and all the conditions for the invariance of the action
under the transformations \cfour\ are satisfied.  Finally, the equivariance
property of $I_r$ written in terms of $({(I_r)^n}_m)$ is
$${f_{a p}}^m  \ {(I_r)^p}_{n} - {f_{a n}}^p  \
{(I_r)^m}_{p}=- {\omega_{a r}}^s {(I_s)^m}_n.
\eqn\ccsix$$

  The commutator of any two left-handed transformations (eqn. \cfour ) closes
on-shell  to left-translations $\delta
X^i=\epsilon^A L^i_A$ and supersymmetry transformations \anine, \cfour\
provided that the following conditions are satisfied:
$$\eqalign {
{{{\CN}} (I_r, I_t)}^p_{mn}&=0
\cr
{(I_r)^n}_p {(I_t)^p}_m &= -\delta _{r t} + {\epsilon_{r t}}^ s
															{(I_s)^n}_m	}
    																										\eqn\cseven$$
and
$${f_{lp}}^a {(I_{(r)})^l}_m {(I_{(r)})^p}_n - {f_{mn}}^a  = {{k^a}_ r}^s
(I_{s})_{mn},																																\eqn\ceight$$
where ${{\CN} (I_r, I_t)}^p_{mn}$ are the components of the Nijenhuis
tensor
$$\eqalign{
{{\CN}(I_r, I_t)^i}_{jk}&= L^i_A\ {{\CN} (I_r, I_t)}^A_{BC}\
L^B_j\  L^C_k
\cr &
= [{(I_r)^h}_{[j} \partial_h{(I_t)^i}_{k]} - \partial_{[j}{(I_r)^h}_{k]}
{(I_t)^i}_{h} + (I_r\rightarrow I_t)]}
        \eqn\cnine$$
restricted on $P$ and $k^a=({{k^a}_ r}^s)$, $a=1,\cdots, {\rm {dim}}{ }LieH$,
are  constant $3\times 3$ matrices (indices in brackets () are not summed
over).

The algebra of left-handed supersymmetry transformations closes
non-linearly to left-translations $\delta X^i=\epsilon^ A L^i_A$ where the
parameter $\epsilon$ of this symmetry includes the currents ${\CJ}_r=I_{rij}
\nabla_+X^i \nabla_+X^j$ of the supersymmetry transformations \cfour, i.e. the
algebra of supersymmetry transformations and left translations of the (4,0)
supersymmetric coset model is  a (super) W-algebra [\zam].   The non-linear
closure of the algebra of the supersymmetry transformations is due to eqn.
\ceight\ and in particular to the fact that some of the $k^a$ matrices are
different from zero.

The conditions given in eqns. \cseven\ and \ceight\ are not all
independent.  In particular, if the components \cseven\ of two the six
${\CN} (I_r, I_s)$ tensors vanish then the same components of the remaining
four tensors  vanish as well.  Similarly, if two of the conditions \ceight\
are satisfied, they imply the third.

To give a geometric interpretation for the conditions \csix -\ceight, we
define the rank three $SO(3)$-vector bundle   ${\CE}$ over the coset space
$G/H$.  This vector bundle is the bundle of the complex structures of
$G/H$ and admits three local sections that obey the algebra of imaginary unit
quaternions.  ${\CE}$ is not always a trivial vector bundle over $G/H$ and it
does not admit global everywhere no-vanishing sections.  We introduce the
frame $e$ and  the canonical connection $\Omega$ of the coset space as in the
previous section. From eqn. \cseven, we can construct a local basis of the
complex structures  given  by  ${I_r^{\nu}}_{\mu}={I_r^{m}}_{n} e^n_{\mu}
e^{\nu}_m$ .  The metric $g$ of $G/H$ is (1,1) with respect to all three
complex
structures.  The latter follows because the matrices $(I_{rmn})$, $r=1,2,3$,
are
antisymmetric.   From eqn. \ccsix, we find that the canonical connection
$\Omega$ of $G/H$ preserves the bundle $\CE$, i.e. the covariant derivative of
a
section of $\CE$ with respect to the canonical connection is again a section
of $\CE$. The equation \ceight\ imposes restrictions on the components of the
curvature tensor of the canonical connection.    The manifold $G/H$ has a
quaternionic structure.  In fact it admits a quaternionic K\"ahler structure
with respect to its canonical connection $\Omega$.   However this quaternionic
K\"ahler structure is not the conventional  one [\ish, \sa] since it is not
with
respect to a Levi-Civita connection.

An interesting class of (4,0) supersymmetric coset models arises whenever the
target manifold $G$ is a {\it {semisimple}} group and $G/H$ is a {\it
{symmetric}} space.  In this case the eqn. \ccsix\ is not independent of eqn.
\ceight\ and the Nijenhuis tensors ${\CN}(I_r,I_t)$ are identically zero.  The
canonical connection of $G/H$ is the Levi-Civita connection of the invariant
metric of $G/H$ and the manifold $G/H$ is a quaternionic K\"ahler manifold
(Wolf space).  The symmetric compact quaternionic K\"ahler manifolds were
classified in Ref [\wolf].


\chapter {Concluding Remarks}
 To  study the conditions for the existence of a $(p,q)$, $p\geq 3,q\geq 1$,
supersymmetric coset model, we start with an off-shell formulation of
the (1,1) supersymmetric coset model as in the construction of (2,2)
one.  A model with (3,q) supersymmetry is in fact
invariant under (4,$q$) supersymmetry transformations.  This is because given
the two complex structures necessary to construct the three left-handed
supersymmetry transformations, we can construct a third one by multiplying
the two complex structures together.  The third complex structure generates
a fourth left-handed supersymmetry transformation.  Thus it is enough to
examine the (4,$q$) supersymmetric coset models.  For this, we introduce
additional $q-1$  right-handed supersymmetry transformations
$$ \eqalign {
\delta_R X^i &= a^s_{+}\  {{J_s}^i}_{j}\ \nabla_-X^{j},
\quad						\delta_R A^a_- = a^s_{+}\  {E_s^a}_b \  F^b_{+-},
\cr
\delta_R A^a_+&=0; \quad
s=1,\cdots,q-1,}
																														\eqn\cten$$
to the four left-handed ones,
where $a^s_+$ are the  parameters of the transformations.   The geometry of
(4,1) supersymmetric coset models is the same as the geometry of the
(4,0) ones.  There are two more models.  These are the (4,2) and the (4,4)
supersymmetric coset models.  For both models, the conditions for the
invariance of the (1,1) action under the additional  left-handed
supersymmetry transformations are the same as those found in the previous
section for the invariance of the (1,0) action under the left-handed (4,0)
supersymmetry transformations. In addition,  the conditions for the
closure of the algebra of additional left-handed supersymmetry
transformations  are given in eqns. \cseven\ and \ceight. In the case of (4,2)
model, the tensor $J=J_r$, $r=1$ , that generates the additional right-handed
supersymmetry transformation is invariant under the adjoint action of the group
$H$ on $G$, and the action \aone\ is invariant under the (4,2) supersymmetry
transformations provided  the tensor $J$ satisfies the condition of eqn.
\atwentythree\ and is antisymmetric.   The algebra of right-handed
supersymmetry transformations closes on-shell subject to conditions satisfied
by the tensor $J$ in section three where the closure properties of the
right-handed supersymmetry transformations of the (2,2) model were examined.
The commutator of left- with right-handed supersymmetry transformations
vanishes on-shell.   Finally in the case of (4,4) model, the tensors $J_r$,
$r=1,2,3$, are {\it equivariant} in a way similar to that of the tensors $I_r$,
$r=1,2,3$.    The (1,1) action is invariant under the additional right-handed
supersymmetry transformations provided that each tensor  $J_r$ satisfies the
condition of eqn. \atwentythree\ and is antisymmetric.  The algebra of
right-handed supersymmetry transformations closes under similar conditions on
$J_r$, $r=1,2,3$ as those described in section five for the tensors $I_r$.  The
commutator of the left-handed with the right-handed supersymmetry
transformations closes on-shell.

{}From the conditions of existence of (4,$q$), $1\leq q \leq 4$, supersymmetric
coset models, it is clear that the  models with (4,1) and (4,2)
supersymmetry are in fact invariant under (4,4) supersymmetry
transformations.   This is because we can construct the tensors $J_r$,
$r=1,2,3$, necessary for the existence of four right-handed supersymmetries
from the tensors $I_r$, $r=1,2,3$ that appear in the left-handed
ones.  Indeed we can    achieve this by  setting
${J_r}^i_j=R^i_A{I_r}^A_B  R^B_j$ where $R$ is the right frame of $G$.
Therefore the \lq\lq independent" supersymmetric coset models are those with
($p$,0) and $(p,p)$, $p=1,2,4$ supersymmetry.

 In conclusion, each $(p,q)$ supersymmetric coset  model has a left-handed and
a
right-handed  Kac-Moody current, a left-handed and a right-handed energy
momentum tensor that generates the translations and first supersymmetry
transformations, and  ${\CJ} _{r\pp}=I_{rij}\nabla_+X^i \nabla_+X^j$, $1\leq r
\leq {p-1}$  and ${\CJ} _{s=}=J_{sij}\nabla_-X^i \nabla_-X^j$, $1\leq s \leq
q-1$ conserved currents that generate the additional supersymmetry
transformations.  In the case of the (4,0) supersymmetric coset model, the
currents ${\CJ} _{r\pp}$  are {\it{covariantly}} conserved. Similarly the
currents ${\CJ} _{r\pp}$ and  ${\CJ} _{s=}$ of the (4,4) supersymmetric coset
model are {\it{covariantly}} conserved as well.  The $(p,q)$ supersymmetric
coset  models exist provided that geometry of the coset space $G/H$ is
restricted appropriately.  In particular, the (2,$q$), $0\leq q \leq 2$  models
exist provided that $G/H$ is a Hermitian manifold with a holomorphic tangent
bundle and the (4,$q$), $0\leq q \leq 4$ models exist provided that $G/H$ is a
quaternionic K\"ahler manifold with respect to the canonical connection of the
coset space.  The algebras of currents of the $(p,q)$ supersymmetric coset
models are superconformal algebras.  Finally, the algebraic closure properties
of the algebra of supersymmetry transformations, the current content and the
geometry of the coset spaces $G/H$ of the (2,2) and (4,0) supersymmetric coset
models indicate that the algebras of currents of these models are closely
related to the N=2 Kazama-Suzuki and N=4 Van Proeyen superconformal algebras
correspondingly.

\noindent{\bf Acknowledgements:} I would like to thank C.M. Hull for advice and
for taking part in the initial stages of this project, and B. Spence
for useful discussions.  This work was supported by the SERC.

\refout

\end